\documentclass[a4paper,11pt]{article}

\usepackage{hyperref}%
\hypersetup{%
  colorlinks=true,%
  linkcolor=blue,%
  citecolor=magenta,%
  urlcolor=black,%
  bookmarksnumbered=true,%
  bookmarkstype=toc,%
  bookmarksopen=false,%
  pdftitle={Scale Invariance Breaking and Discrete Phase Invariance in Few-Body Problems},%
  pdfkeywords={discrete phase invariance; two-body problem; three-body problem},%
  pdfauthor={Satoshi Ohya}}%

\usepackage[margin=1in]{geometry}%

\usepackage[tt=false,sb]{libertine}%
\usepackage[T1]{fontenc}%
\usepackage{libertinust1math}%
\usepackage[cal=stix,scr=boondoxo,bb=boondox]{mathalpha}%
\usepackage{bm}%
\usepackage{mathtools}%
\DeclareMathOperator{\e}{e}%
\DeclareMathOperator*{\re}{Re}%
\DeclareMathOperator*{\im}{Im}%
\DeclareMathOperator{\sgn}{sgn}%
\DeclareMathOperator*{\Res}{Res}%
\allowdisplaybreaks[1]%

\usepackage{tikz}

\usepackage[font=small,labelfont=bf]{caption}%

\usepackage[title]{appendix}

\usepackage[backend=biber,style=phys,biblabel=brackets,block=ragged,eprint=true,url=true]{biblatex}%
\title{\Large\bfseries\boldmath Scale Invariance Breaking and Discrete Phase Invariance\\ in Few-Body Problems}%
\author{\normalsize Satoshi Ohya\\[1em]
  \small\itshape Institute of Quantum Science, Nihon University,\\
  \small\itshape Kanda-Surugadai 1-8-14, Chiyoda, Tokyo 101-8308, Japan\\[1ex]
  \small\ttfamily ohya.satoshi@nihon-u.ac.jp}%
\date{\small(Dated: \today)}%

\begin{document}
\maketitle%
\flushbottom%

\begin{abstract}
  Scale invariance in quantum mechanics can be broken in several
  ways. A well-known example is the breakdown of continuous scale
  invariance to discrete scale invariance, whose typical realization
  is the Efimov effect of three-body problems. Here we discuss yet
  another discrete symmetry to which continuous scale invariance can
  be broken: \textit{discrete phase invariance}. We first revisit the
  one-body problem on the half line in the presence of an
  inverse-square potential---the simplest example of nontrivial
  scale-invariant quantum mechanics---and show that continuous scale
  invariance can be broken to discrete phase invariance in a small
  window of coupling constant. We also show that discrete phase
  invariance manifests itself as circularly distributed simple poles
  on Riemann sheets of the S-matrix. We then present three examples of
  few-body problems that exhibit discrete phase invariance. These
  examples are the one-body Aharonov-Bohm problem, a two-body problem
  of nonidentical particles in two dimensions, and a three-body
  problem of nonidentical particles in one dimension, all of which
  contain a codimension-two ``magnetic'' flux in configuration spaces.
\end{abstract}

\newpage
\section{Introduction}
\label{section:1}
One of the most surprising pattern of scale invariance breaking is the
breakdown of continuous scale invariance to discrete scale invariance
(or log-periodicity \cite{Sornette:1997pb}). As is well known,
discrete scale invariance implies the existence of a geometric
sequence of bound-state poles in S-matrix elements, whose typical
realization is the Efimov effect \cite{Efimov:1970zz} of three
identical bosons in three dimensions; see, e.g., \cite{Naidon:2016dpf}
for review. The purpose of this paper is to discuss yet another
breaking pattern of scale invariance in few-body problems: breakdown
of continuous scale invariance to \textit{discrete phase invariance}
(or imaginary log-periodicity)---a complexified version of discrete
scale invariance (or log-periodicity along the imaginary
direction). As we will see below, discrete phase invariance implies
the existence of circularly distributed simple poles in S-matrix
elements, which may cause particular resonance-like effects in
particle scatterings.

The rest of the paper is organized as follows. We first revisit the
simplest problem that exhibits scale invariance breaking: the
one-dimensional one-body problem under the inverse-square potential
$V(r)=\lambda/r^{2}$. This problem enjoys continuous scale invariance
if the coupling constant $\lambda$ is larger than an upper critical
value $\lambda_{\ast\ast}$. It is well known, however, that this
continuous scale invariance is broken to discrete scale invariance if
the coupling constant $\lambda$ is smaller than a lower critical value
$\lambda_{\ast}$ \cite{Case:1950an}. Section \ref{section:2} considers
the case in which the coupling constant lies in the intermediate
window $\lambda_{\ast}<\lambda<\lambda_{\ast\ast}$. In this case, we
show that continuous scale invariance is broken to discrete phase
invariance. In particular, we derive the S-matrix (reflection
amplitude off the boundary) exactly and show that it is indeed
log-periodic along the imaginary axis and possesses circularly
distributed simple poles in the complex momentum space. Section
\ref{section:3} presents several few-body examples that boil down to
the inverse-square-potential problem in the intermediate
window. Examples include a two-dimensional two-body problem and a
one-dimensional three-body problem in the presence of a
codimension-two ``magnetic'' flux in the many-body configuration
spaces. We conclude in section \ref{section:4}. Appendix
\ref{appendix:A} proves the orthonormality and completeness of energy
eigenfunctions for the inverse-square-potential problem in the
intermediate window.

\section{Inverse-square potential in the intermediate window}
\label{section:2}
In this section, we revisit the problem of inverse-square potential in
the intermediate window, where continuous scale invariance is
broken. This problem has been studied over sixty years in various
contexts of physics, including self-adjoint extension
\cite{Meetz:1964}, scale anomaly \cite{Camblong:2001zt},
renormalization \cite{Bouaziz:2014wxa}, and few-body resonances
\cite{Nishida:2007mr,Pricoupenko:2023rok}. To the best of our
knowledge, however, its breaking pattern of scale invariance has not
been elucidated before. In particular, the appearance of circularly
distributed simple poles in S-matrix has not been appreciated in the
literature. In the following, we first discuss (without proof) the
basic idea of scale invariance breaking in the problem by using a
scaling law. Computational details are relegated to the subsequent
subsections. For simplicity, throughout this section we will use the
units $\hbar=2m=1$, where $m$ is the particle mass.

Consider the following one-body Hamiltonian on the half line
$\mathbb{R}_{+}=\{r:r>0\}$:
\begin{align}
  H=-\frac{d^{2}}{dr^{2}}+\frac{\lambda}{r^{2}},\label{eq:1}
\end{align}
where $\lambda\in(-\infty,\infty)$ is a dimensionless coupling
constant. Under the scale transformation $r\mapsto\e^{t}r$, the
Hamiltonian \eqref{eq:1} is transformed as $H\mapsto\e^{-2t}H$, where
$t$ is an arbitrary constant. Hence, if $\psi_{E}(r)$ is an energy
eigenfunction satisfying $H\psi_{E}(r)=E\psi_{E}(r)$, it also
satisfies $\e^{-2t}H\psi_{E}(\e^{t}r)=E\psi_{E}(\e^{t}r)$; that is,
$H\psi_{E}(\e^{t}r)=\e^{2t}E\psi_{E}(\e^{t}r)$. Hence,
$\psi_{\e^{2t}E}(r)$ must be proportional to $\psi_{E}(\e^{t}r)$:
\begin{align}
  \psi_{\e^{2t}E}(r)\propto\psi_{E}(\e^{t}r).\label{eq:2}
\end{align}
This equation---the scaling law for the energy eigenfunction---reduces
to a scaling law for the S-matrix. To see this, consider the
asymptotic behavior of the energy eigenfunction
\begin{align}
  \psi_{E}(r)\propto\e^{-i\sqrt{E}r}+S(E)\e^{+i\sqrt{E}r}\quad\text{as}\quad r\to\infty,\label{eq:3}
\end{align}
where $S(E)$ is the S-matrix (reflection amplitude off the
boundary). Substituting this into \eqref{eq:2} and comparing both
sides, we get
\begin{align}
  S(\e^{2t}E)=S(E).\label{eq:4}
\end{align}
This is the scaling law for the S-matrix.

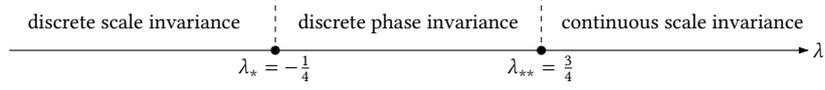
\begin{figure}[t]
  \centering%
\xdefinecolor{rgb_000000}{rgb}{0,0,0}%
\begin{tikzpicture}
\pgfsetlinewidth{0.4pt}
\useasboundingbox (0cm,0cm) rectangle (10.85cm,1.225cm);
\draw (0.175cm,0.525cm)--(10.675cm,0.525cm);
\pgfsetfillcolor{rgb_000000}
\draw [fill](10.5696cm,0.525cm)--(10.5696cm,0.489854cm)--
  (10.675cm,0.525cm)--(10.5696cm,0.560146cm)--(10.5696cm,0.525cm)--cycle;
\filldraw[color=rgb_000000] (3.675cm,0.525cm) circle(0.052719cm);
\filldraw[color=rgb_000000] (7.175cm,0.525cm) circle(0.052719cm);
\draw (3.675cm,0.525cm)--(3.675cm,0.56cm);
\draw (3.675cm,0.63cm)--(3.675cm,0.665cm);
\draw (3.675cm,0.665cm)--(3.675cm,0.7cm);
\draw (3.675cm,0.77cm)--(3.675cm,0.805cm);
\draw (3.675cm,0.805cm)--(3.675cm,0.84cm);
\draw (3.675cm,0.91cm)--(3.675cm,0.945cm);
\draw (3.675cm,0.945cm)--(3.675cm,0.98cm);
\draw (3.675cm,1.05cm)--(3.675cm,1.085cm);
\draw (3.675cm,1.085cm)--(3.675cm,1.12cm);
\draw (3.675cm,1.19cm)--(3.675cm,1.225cm);
\draw (7.175cm,0.525cm)--(7.175cm,0.56cm);
\draw (7.175cm,0.63cm)--(7.175cm,0.665cm);
\draw (7.175cm,0.665cm)--(7.175cm,0.7cm);
\draw (7.175cm,0.77cm)--(7.175cm,0.805cm);
\draw (7.175cm,0.805cm)--(7.175cm,0.84cm);
\draw (7.175cm,0.91cm)--(7.175cm,0.945cm);
\draw (7.175cm,0.945cm)--(7.175cm,0.98cm);
\draw (7.175cm,1.05cm)--(7.175cm,1.085cm);
\draw (7.175cm,1.085cm)--(7.175cm,1.12cm);
\draw (7.175cm,1.19cm)--(7.175cm,1.225cm);
\pgftext[at={\pgfpoint{10.7453cm}{0.525cm}}] {\makebox(0,0)[l]{\hbox{\color{rgb_000000}\scriptsize $\lambda$}}}
\pgftext[at={\pgfpoint{3.675cm}{0.454708cm}}] {\makebox(0,0)[t]{\hbox{\color{rgb_000000}\scriptsize $\lambda_{\ast}=-\frac{1}{4}$}}}
\pgftext[at={\pgfpoint{7.175cm}{0.454708cm}}] {\makebox(0,0)[t]{\hbox{\color{rgb_000000}\scriptsize $\lambda_{\ast\ast}=\frac{3}{4}$}}}
\pgftext[at={\pgfpoint{8.925cm}{0.875cm}}] {\makebox(0,0)[c]{\hbox{\color{rgb_000000}\scriptsize \phantom{p} continuous scale invariance}}}
\pgftext[at={\pgfpoint{1.925cm}{0.875cm}}] {\makebox(0,0)[c]{\hbox{\color{rgb_000000}\scriptsize discrete scale invariance \phantom{p}}}}
\pgftext[at={\pgfpoint{5.425cm}{0.875cm}}] {\makebox(0,0)[c]{\hbox{\color{rgb_000000}\scriptsize discrete phase invariance}}}
\end{tikzpicture}
  \caption{``Phases'' of the inverse-square-potential problem. There
    exist two critical values of the coupling constant $\lambda$: the
    upper critical value $\lambda_{\ast\ast}=3/4$ and the lower
    critical value $\lambda_{\ast}=-1/4$. These two values are
    determined by the boundary behavior of energy eigenfunction. Near
    the boundary, solutions to the eigenvalue equation
    $H\psi_{E}(r)=E\psi_{E}(r)$ are well approximated by the
    zero-energy solutions given by $r^{1/2\pm\sqrt{\lambda+1/4}}$. If
    these solutions are square integrable near the boundary, the
    general solution behaves as a linear combination
    $Ar^{1/2+\sqrt{\lambda+1/4}}+Br^{1/2-\sqrt{\lambda+1/4}}$ near the
    boundary. But $r^{1/2+\sqrt{\lambda+1/4}}$ and
    $r^{1/2-\sqrt{\lambda+1/4}}$ have different scaling dimensions,
    such a linear combination must introduce a scale, which breaks
    continuous scale invariance. On the other hand, if one of these
    solutions is non-square integrable even in a small interval
    $0<r<\varepsilon$, there is no chance to introduce a scale so that
    continuous scale invariance remains intact. For
    $\lambda>\lambda_{\ast\ast}=3/4$, the solution
    $r^{1/2-\sqrt{\lambda+1/4}}$ is non-square integrable near the
    boundary and hence continuous scale invariance is never
    broken. This is the continuous-scale-invariant phase above the
    upper critical value. For $\lambda<\lambda_{\ast\ast}$, on the
    other hand, both the solutions are square integrable near the
    boundary and hence continuous scale invariance is broken. But for
    $\lambda=\lambda_{\ast}<-1/4$, the linear combination
    $Ar^{1/2+i\sqrt{-1/4-\lambda}}+Br^{1/2-i\sqrt{-1/4-\lambda}}$
    becomes invariant (up to an overall factor) under the discrete
    scale transformation $r\mapsto\e^{t}r$ with
    $t=n\pi/\sqrt{-1/4-\lambda}$, where $n$ is an integer. This is the
    discrete-scale-invariant phase below the lower critical value
    $\lambda_{\ast}=-1/4$. Furthermore, for
    $\lambda_{\ast}<\lambda<\lambda_{\ast\ast}$, the linear
    combination
    $Ar^{1/2+\sqrt{\lambda+1/4}}+Br^{1/2-\sqrt{\lambda+1/4}}$ becomes
    invariant (up to an overall factor) under the discrete phase
    transformation $r\mapsto\e^{t}r$ with
    $t=in\pi/\sqrt{\lambda+1/4}$. This is the discrete-phase-invariant
    phase in the intermediate window.}
  \label{figure:1}
\end{figure}

Now, if this equation holds for arbitrary $t$, then the S-matrix must
be an $E$-independent constant, because there is no other solution to
the equation \eqref{eq:4}. In the region below the upper critical
value $\lambda_{\ast\ast}$, however, the scaling law \eqref{eq:4} does
not hold for arbitrary $t$ in general; see figure \ref{figure:1} and
its caption for an explanation of this point. In the following, we
discuss breakdown of the scaling law \eqref{eq:4} and its consequence
under the assumption that the S-matrix has a bound-state pole at
$E=E_{0}$:
\begin{align}
  S(E)=\frac{\text{const.}}{E-E_{0}}+O(1)\quad\text{as}\quad E\to E_{0}.\label{eq:5}
\end{align}
This assumption is of course justified by calculating the S-matrix
explicitly.

Let us first consider the region
$\lambda\in(-\infty,\lambda_{\ast})$. In this region, continuous scale
invariance is broken to discrete scale invariance; that is, the
scaling law \eqref{eq:4} holds only for some quantized $t$. An
explicit calculation shows that the S-matrix fulfills the following
equality for arbitrary integer $n$:
\begin{align}
  S(\e^{\frac{2n\pi}{\sqrt{\lambda_{\ast}-\lambda}}}E)=S(E)\quad\text{for}\quad\lambda\in(-\infty,\lambda_{\ast}).\label{eq:6}
\end{align}
An immediate consequence of this discrete scaling law is the
appearance of geometric series of bound-state poles in the
S-matrix. Combining \eqref{eq:6} with \eqref{eq:5}, we get
\begin{align}
  S(E)
  =S(\e^{\frac{2n\pi}{\sqrt{\lambda_{\ast}-\lambda}}}E)
  =\frac{\text{const.}}{\e^{\frac{2n\pi}{\sqrt{\lambda_{\ast}-\lambda}}}E-E_{0}}+O(1)
  \quad\text{as}\quad \e^{\frac{2n\pi}{\sqrt{\lambda_{\ast}-\lambda}}}E\to E_{0},\label{eq:7}
\end{align}
or, equivalently,
\begin{align}
  S(E)=\frac{\text{const.}\times\e^{-\frac{2n\pi}{\sqrt{\lambda_{\ast}-\lambda}}}}{E-E_{0}\e^{-\frac{2n\pi}{\sqrt{\lambda_{\ast}-\lambda}}}}+O(1)\quad\text{as}\quad E\to E_{0}\e^{-\frac{2n\pi}{\sqrt{\lambda_{\ast}-\lambda}}}.\label{eq:8}
\end{align}
Hence, if there exists a simple pole at $E=E_{0}$, there in fact exist
infinitely many simple poles at $E=E_{n}$, where
\begin{align}
  E_{n}=E_{0}\exp\left(-\frac{2n\pi}{\sqrt{\lambda_{\ast}-\lambda}}\right)\quad\text{for}\quad\lambda\in(-\infty,\lambda_{\ast}).\label{eq:9}
\end{align}
This is the geometric series of bound-state energies
\cite{Case:1950an}.

Let us next consider the intermediate window
$\lambda\in(\lambda_{\ast},\lambda_{\ast\ast})$. In this window,
continuous scale invariance is broken to discrete phase invariance;
that is, the scaling law \eqref{eq:4} holds for some quantized
imaginary $t$. This can be easily guessed by noticing that, for
$\lambda>\lambda_{\ast}$, the square root
$\sqrt{\lambda_{\ast}-\lambda}$ becomes the pure imaginary
$i\sqrt{\lambda-\lambda_{\ast}}$. Correspondingly, the discrete
scaling law \eqref{eq:6} is replaced by
\begin{align}
  S(\e^{-i\frac{2n\pi}{\sqrt{\lambda-\lambda_{\ast}}}}E)=S(E)\quad\text{for}\quad\lambda\in(\lambda_{\ast},\lambda_{\ast\ast}).\label{eq:10}
\end{align}
Combining this with \eqref{eq:5}, we get
\begin{align}
  S(E)=\frac{\text{const.}\times\e^{i\frac{2n\pi}{\sqrt{\lambda-\lambda_{\ast}}}}}{E-E_{0}\e^{i\frac{2n\pi}{\sqrt{\lambda-\lambda_{\ast}}}}}+O(1)\quad\text{as}\quad E\to\e^{i\frac{2n\pi}{\sqrt{\lambda_{\ast}-\lambda}}}E_{0}.\label{eq:11}
\end{align}
Hence, in the intermediate window, the S-matrix has complex poles at
$E=E_{n}$, where
\begin{align}
  E_{n}
  =E_{0}\e^{i\frac{2n\pi}{\sqrt{\lambda_{\ast}-\lambda}}}
  =E_{0}\cos\left(\frac{2n\pi}{\sqrt{\lambda-\lambda_{\ast}}}\right)+iE_{0}\sin\left(\frac{2n\pi}{\sqrt{\lambda-\lambda_{\ast}}}\right)
  \quad\text{for}\quad\lambda\in(\lambda_{\ast},\lambda_{\ast\ast}).\label{eq:12}
\end{align}
These are the circularly distributed simple poles along a circle of
radius $|E_{0}|$ in the complex $E$-plane. Notice that, since
$\sqrt{\lambda-\lambda_{\ast}}<1$ for
$\lambda\in(\lambda_{\ast},\lambda_{\ast\ast})$, the argument of the
complex number
$E_{n}=E_{0}\exp(i\frac{2n\pi}{\sqrt{\lambda-\lambda_{\ast}}})$ for
$n\neq0$ is not in the interval $[0,2\pi)$; that is, all the complex
poles lie in the unphysical sheets. Note also that the number of
complex poles is finite (infinite) if $\sqrt{\lambda-\lambda_{\ast}}$
is a rational (irrational) number.

We emphasize that \eqref{eq:9} is well known but \eqref{eq:12} is
not. In the rest of this section, we derive the results
\eqref{eq:10}--\eqref{eq:12} by solving the eigenvalue equation
$H\psi_{E}(r)=E\psi_{E}(r)$ exactly.

\subsection{Boundary condition}
\label{section:2.1}
In order to solve the eigenvalue equation, we first have to specify
the boundary condition at $r=0$. The boundary condition must be the
one that guarantees self-adjointness of the Hamiltonian
\eqref{eq:1}. It is, however, well known that \eqref{eq:1} is not
essentially self-adjoint in the intermediate window; rather, it has a
one-parameter family of self-adjoint extensions. Namely, the most
general boundary condition at $r=0$ in the intermediate window
contains a single real parameter. In what follows, following the idea
of probability-conservation requirement as an alternative to
self-adjoint extension \cite{Fulop:1999pf}, we give a simple
derivation for the one-parameter family of boundary conditions given
by \cite{Ohya:2010zm} without dwelling too much on mathematics. For
more details on the mathematical aspect of self-adjoint extension in
the inverse-square-potential problem, we refer to \cite{Meetz:1964}.

The key to our derivation is a factorization of the Hamiltonian. As is
well known in the context of supersymmetric quantum mechanics (see,
e.g., Section 2 of \cite{Cooper:1994eh}), the Hamiltonian \eqref{eq:1}
can be factorized as $H=A^{-}A^{+}$, where $A^{\pm}$ are first-order
differential operators given by
$A^{+}=\psi_{0}(r)\frac{d}{dr}\psi_{0}(r)^{-1}$ and
$A^{-}=-\psi_{0}(r)^{-1}\frac{d}{dr}\psi_{0}(r)$. Here $\psi_{0}(r)$
is a zero-energy solution satisfying $H\psi_{0}(r)=0$. In the present
problem, there are two such zero-energy solutions,
$r^{1/2+\sqrt{\lambda-\lambda_{\ast}}}$ and
$r^{1/2-\sqrt{\lambda-\lambda_{\ast}}}$, the former always vanishes at
the boundary whereas the latter can blow up as $r\to0$. By choosing
the latter, we get the following factorization of \eqref{eq:1}:
\begin{align}
  H=-\frac{1}{r^{\frac{1}{2}-\nu}}\frac{d}{dr}r^{1-2\nu}\frac{d}{dr}\frac{1}{r^{\frac{1}{2}-\nu}},\quad\text{where}\quad\nu\coloneq\sqrt{\lambda-\lambda_{\ast}}\in(0,1).\label{eq:13}
\end{align}
As we will see soon, the factor $1/r^{1/2-\nu}$ kills the singular
part of wavefunctions and makes the boundary condition well defined at
$r=0$.

Now we consider the probability conservation. In general, the
probability conservation refers to the time-independence of the
following quantity (total probability):
\begin{align}
  Q=\int_{0}^{\infty}\!\!\!dr\,\overline{\psi}\psi,\label{eq:14}
\end{align}
where $\psi$ is an arbitrary wavefunction that satisfies the
time-dependent Schr\"{o}dinger equation
$i\partial\psi/\partial t=H\psi$ and the overline
($\overline{\phantom{m}}$) stands for the complex conjugate. Without
boundary, the conservation of $Q$ is almost trivially satisfied (under
the assumption that $\psi$ vanishes at infinity). In the presence of
boundary, however, the conservation of $Q$ becomes a nontrivial
problem. In fact, a straightforward calculation gives
\begin{align}
  \frac{d}{dt}Q
  &=\int_{0}^{\infty}\!\!\!dr\,\frac{\partial}{\partial t}(\overline{\psi}\psi)\nonumber\\
  &=\int_{0}^{\infty}\!\!\!dr\left(\overline{\psi}\frac{\partial\psi}{\partial t}+\frac{\partial\overline{\psi}}{\partial t}\psi\right)\nonumber\\
  &=\frac{1}{i}\int_{0}^{\infty}\!\!\!dr\left[\overline{\psi}H\psi-(H\overline{\psi})\psi\right]\nonumber\\
  &=\frac{1}{i}\int_{0}^{\infty}\!\!\!dr\left[-\frac{\overline{\psi}}{r^{\frac{1}{2}-\nu}}\frac{\partial}{\partial r}\left(r^{1-2\nu}\frac{\partial}{\partial r}\frac{\psi}{r^{\frac{1}{2}-\nu}}\right)+\left(\frac{\partial}{\partial r}\left(r^{1-2\nu}\frac{\partial}{\partial r}\frac{\overline{\psi}}{r^{\frac{1}{2}-\nu}}\right)\right)\frac{\psi}{r^{\frac{1}{2}-\nu}}\right]\nonumber\\
  &=\frac{1}{i}\int_{0}^{\infty}\!\!\!dr\,\frac{\partial}{\partial r}\left[-\frac{\overline{\psi}}{r^{\frac{1}{2}-\nu}}\left(r^{1-2\nu}\frac{\partial}{\partial r}\frac{\psi}{r^{\frac{1}{2}-\nu}}\right)+\left(r^{1-2\nu}\frac{\partial}{\partial r}\frac{\overline{\psi}}{r^{\frac{1}{2}-\nu}}\right)\frac{\psi}{r^{\frac{1}{2}-\nu}}\right]\nonumber\\
  &=\frac{1}{i}\lim_{r\to0}\left[\frac{\overline{\psi}}{r^{\frac{1}{2}-\nu}}\left(r^{1-2\nu}\frac{\partial}{\partial r}\frac{\psi}{r^{\frac{1}{2}-\nu}}\right)-\left(r^{1-2\nu}\frac{\partial}{\partial r}\frac{\overline{\psi}}{r^{\frac{1}{2}-\nu}}\right)\frac{\psi}{r^{\frac{1}{2}-\nu}}\right],\label{eq:15}
\end{align}
where in the third line we have used the time-dependent
Schr\"{o}dinger equation $\partial\psi/\partial t=(1/i)H\psi$ and its
complex conjugate
$\partial\overline{\psi}/\partial t=-(1/i)H\overline{\psi}$ and in the
fourth line we have used the factorization \eqref{eq:13}. In the last
line we have assumed that the wavefunction $\psi$ rapidly vanishes as
$r\to\infty$. Hence, in order to guarantee the probability
conservation $dQ/dt=0$, $\psi$ must satisfy
\begin{align}
  \frac{\overline{\psi}}{r^{\frac{1}{2}-\nu}}\left(r^{1-2\nu}\frac{\partial}{\partial r}\frac{\psi}{r^{\frac{1}{2}-\nu}}\right)-\left(r^{1-2\nu}\frac{\partial}{\partial r}\frac{\overline{\psi}}{r^{\frac{1}{2}-\nu}}\right)\frac{\psi}{r^{\frac{1}{2}-\nu}}=0\label{eq:16}
\end{align}
at the boundary $r=0$.

Now we wish to solve this equation. To this end, let us consider the
following simplified problem:
\begin{align}
  \overline{z}w-\overline{w}z=0,\label{eq:17}
\end{align}
where $z$ and $w$ correspond to $\psi/r^{1/2-\nu}$ and
$r^{1-2\nu}(\partial/\partial r)(\psi/r^{1/2-\nu})$, respectively,
both of which are well-defined at $r=0$. Notice that \eqref{eq:17} is
equivalent to the following equality:
\begin{align}
  |z-ig_{0}w|^{2}=|z+ig_{0}w|^{2},\label{eq:18}
\end{align}
where $g_{0}$ is an arbitrary nonvanishing real. Indeed, by using
$|z\mp ig_{0}w|^{2}=|z|^{2}+g_{0}^{2}|w|^{2}\pm
ig_{0}(\overline{z}w-\overline{w}z)$, one immediately sees that
\eqref{eq:18} is equivalent to \eqref{eq:17} provided $g_{0}$ is
nonvanishing. Eq.~\eqref{eq:18} says that two distinct complex numbers
$z-ig_{0}w$ and $z+ig_{0}w$ have the same absolute value. Hence they
must be related by a $U(1)$ transformation. Thus we can write
\begin{align}
  z-ig_{0}w=\e^{i\varphi}(z+ig_{0}w),\quad\e^{i\varphi}\in U(1)\label{eq:19}
\end{align}
with some angle parameter $\varphi$. This is the $U(1)$ family of
linearized relations that solve the quadratic equation
\eqref{eq:17}. Putting $z=\psi/r^{1/2-\nu}$ and
$w=r^{1-2\nu}(\partial/\partial r)(\psi/r^{1/2-\nu})$ into this, we
get
\begin{align}
  \frac{\psi}{r^{\frac{1}{2}-\nu}}+gr^{1-2\nu}\frac{\partial}{\partial r}\left(\frac{\psi}{r^{\frac{1}{2}-\nu}}\right)=0\quad\text{at}\quad r=0,\label{eq:20}
\end{align}
where $g\coloneq g_{0}\cot(\varphi/2)$. This is the most general
boundary condition that guarantees the probability conservation
$dQ/dt=0$ in the intermediate window. Notice that $g$ is a
dimensionful parameter with length dimension $2\nu$. As we will see
below, this dimensionful parameter breaks continuous scale invariance
and serves as a source of energy scale $E_{0}$ in the previous
subsection.

In the subsequent subsections, we will solve the eigenvalue equation
$H\psi=E\psi$ under the boundary condition \eqref{eq:20}. To do this,
it is convenient to introduce the following two linearly independent
functions $f_{\nu}$ and $\overline{f_{\nu}}$ (see, e.g., Eq.~(2.14) of
\cite{Meetz:1964}):
\begin{subequations}
  \begin{align}
    f_{\nu}(z)&=\sqrt{\frac{+\pi iz}{2}}\e^{+i\frac{\nu}{2}\pi}H^{(1)}_{\nu}(z),\label{eq:21a}\\
    \overline{f_{\nu}(\overline{z})}&=\sqrt{\frac{-\pi iz}{2}}\e^{-i\frac{\nu}{2}\pi}H^{(2)}_{\nu}(z),\label{eq:21b}
  \end{align}
\end{subequations}
where $H^{(1)}_{\nu}$ and $H^{(2)}_{\nu}$ are the Hankel functions of
the first and second kind, respectively. These two functions are
linearly independent solutions to the differential equation
$-\frac{d^{2}y}{dz^{2}}+\frac{\nu^{2}-1/4}{z^{2}}y=y$ and satisfy the
following properties:\footnote{Eq.~\eqref{eq:22a} is just the Schwarz
  reflection principle for $f_{\nu}(z)$. To see this, notice that
  $f_{\nu}(z)$ can be real for positive imaginary $z$. Note also that
  the reflection of $z$ with respect to the imaginary axis is
  $-\Bar{z}$. Hence the Schwarz reflection principle gives
  $f_{\nu}(z)=\overline{f_{\nu}(-\overline{z})}$.}
\begin{subequations}
  \begin{align}
    &\overline{f_{\nu}(z)}=f_{\nu}(-\overline{z}),\label{eq:22a}\\
    &f_{\nu}(z)\to\e^{iz}\quad\text{as}\quad|z|\to\infty,\label{eq:22b}\\
    &f_{\nu}(z)\to(-iz)^{\frac{1}{2}-\nu}A_{\nu}-(-iz)^{\frac{1}{2}+\nu}B_{\nu}\quad\text{as}\quad|z|\to0,\label{eq:22c}
  \end{align}
\end{subequations}
where
\begin{align}
  A_{\nu}=\frac{\sqrt{\pi}}{2^{\frac{1}{2}-\nu}\Gamma(1-\nu)\sin(\nu\pi)}
  \quad\text{and}\quad
  B_{\nu}=\frac{\sqrt{\pi}}{2^{\frac{1}{2}+\nu}\Gamma(1+\nu)\sin(\nu\pi)}.\label{eq:23}
\end{align}
We note that $f_{\nu}$ and $\overline{f_{\nu}}$ play the roles of
Jost-like solutions in the scattering theory of inverse-square
potential.

\subsection{Bound-state problem}
\label{section:2.2}
Let us first consider the case $E<0$ and write $E=-\kappa^{2}$, where
$\kappa>0$. We are interested in a square-integrable solution
$\psi_{\kappa}(r)$ whose asymptotic behavior as $r\to\infty$ is
$\psi_{\kappa}(r)\to N_{\kappa}\e^{-\kappa r}$, where $N_{\kappa}$ is
a normalization constant. By using \eqref{eq:21a}, such a solution is
given by
\begin{align}
  \psi_{\kappa}(r)=N_{\kappa}f_{\nu}(i\kappa r).\label{eq:24}
\end{align}
It then follows from \eqref{eq:22c} that this solution has the
following boundary behavior:
\begin{align}
  \psi_{\kappa}(r)\to N_{\kappa}\left((\kappa r)^{\frac{1}{2}-\nu}A_{\nu}-(\kappa r)^{\frac{1}{2}+\nu}B_{\nu}\right)\quad\text{as}\quad r\to0.\label{eq:25}
\end{align}
Substituting this into the boundary condition \eqref{eq:20}, we get
\begin{align}
  \kappa^{\frac{1}{2}-\nu}A_{\nu}-2\nu g\kappa^{\frac{1}{2}+\nu}B_{\nu}=0,\label{eq:26}
\end{align}
or, equivalently,
\begin{align}
  \kappa^{2\nu}=\sgn(g)\kappa_{0}^{2\nu},\quad\text{where}\quad\kappa_{0}\coloneq\left(\frac{A_{\nu}}{2\nu B_{\nu}}\frac{1}{|g|}\right)^{\frac{1}{2\nu}}>0.\label{eq:27}
\end{align}
Here $\sgn(\ast)$ stands for the sign function defined by $\sgn(g)=1$
for $g>0$ and $\sgn(g)=-1$ for $g<0$. Eq.~\eqref{eq:27} gives the
quantization condition for $\kappa$. For $g>0$, this equation has a
single positive solution given by
$\kappa=\kappa_{0}$. Correspondingly, there exists a single bound
state with the energy eigenvalue $E_{0}=-\kappa_{0}^{2}$ for
$g>0$. Notice that the quantization condition \eqref{eq:27} possesses
complex solutions as well. For instance, for $g>0$, the equation
$\kappa^{2\nu}=\kappa_{0}^{2\nu}$ has complex solutions
$\kappa=\kappa_{0}\e^{i\frac{2n\pi}{2\nu}}$ for arbitrary integer
$n$. For $g<0$, on the other hand, the equation
$\kappa^{2\nu}=-\kappa_{0}^{2\nu}$ has complex solutions
$\kappa=\kappa_{0}\e^{i\frac{(2n+1)\pi}{2\nu}}$. As we will see in
\eqref{eq:34d} and \eqref{eq:34e}, these complex solutions appear as
simple poles on the higher Riemann sheets of the S-matrix.

Finally, we note that $N_{\kappa}$ is determined by the normalization
condition $\int_{0}^{\infty}\!dr\,|\psi_{\kappa}(r)|^{2}=1$. The
result is (see appendix \ref{appendix:A})
\begin{align}
  |N_{\kappa}|=\sqrt{\frac{\kappa\sin(\nu\pi)}{\nu}}.\label{eq:28}
\end{align}
As we will see soon, this normalization constant appears as the
residue of the S-matrix at the bound-state pole.

\subsection{Scattering problem}
\label{section:2.3}
\begin{figure}[t]
  \centering%
  \input{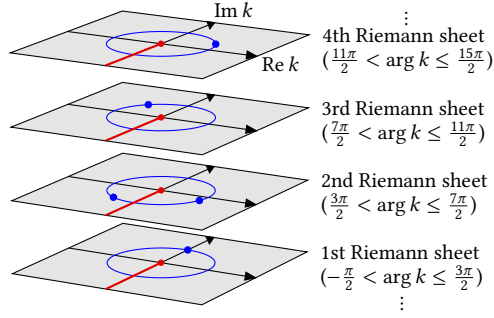}
  \caption{Typical simple-pole distribution on the Riemann surface of
    the S-matrix \eqref{eq:32} for $g>0$. Blue dots represent the
    simple poles, which are distributed along the blue circle of
    radius $\kappa_{0}$. Red dots represent the branch point at
    $k=0$. Branch cuts are chosen along the negative imaginary axis
    and represented by red lines. The whole Riemann surface is given
    by gluing adjacent sheets along the branch cut. In this figure,
    $\nu$ is chosen as the irrational number $\nu=1/\sqrt{2}$ so that
    the simple poles are distributed at
    $k=i\kappa_{0}\e^{in\pi/\nu}=\kappa_{0}\e^{i(1/2+\sqrt{2}n)\pi}$.}
  \label{figure:2}
\end{figure}

Let us next consider the case $E>0$ and write $E=k^{2}$, where
$k>0$. In this case, we are interested in a solution $\psi_{k}(r)$
whose asymptotic behavior as $r\to\infty$ is
$\psi_{k}(r)\to\e^{-ikr}+S_{\nu}(k)\e^{ikr}$, where $S_{\nu}(k)$ is
the S-matrix. By using \eqref{eq:22a} and \eqref{eq:22b}, such a
solution is given by
\begin{align}
  \psi_{k}(r)=f_{\nu}(-kr)+S_{\nu}(k)f_{\nu}(kr).\label{eq:29}
\end{align}
It then follows from \eqref{eq:22c} that this solution has the
following boundary behavior:
\begin{align}
  \psi_{k}(r)\to\left((+ikr)^{\frac{1}{2}-\nu}+S_{\nu}(k)(-ikr)^{\frac{1}{2}-\nu}\right)A_{\nu}-\left((+ikr)^{\frac{1}{2}+\nu}+S_{\nu}(k)(-ikr)^{\frac{1}{2}+\nu}\right)B_{\nu}\quad\text{as}\quad r\to0.\label{eq:30}
\end{align}
Substituting this into the boundary condition \eqref{eq:20}, we get
\begin{align}
  A_{\nu}\left((+ik)^{\frac{1}{2}-\nu}+S_{\nu}(k)(-ik)^{\frac{1}{2}-\nu}\right)-2\nu gB_{\nu}\left((+ik)^{\frac{1}{2}+\nu}+S_{\nu}(k)(-ik)^{\frac{1}{2}+\nu}\right)=0.\label{eq:31}
\end{align}
By solving this in terms of $S_{\nu}(k)$ and using the relation
$2\nu g B_{\nu}/A_{\nu}=\sgn(g)\kappa_{0}^{-2\nu}$, we get
\begin{align}
  S_{\nu}(k)=-\frac{(+ik/\kappa_{0})^{\frac{1}{2}-\nu}-\sgn(g)(+ik/\kappa_{0})^{\frac{1}{2}+\nu}}{(-ik/\kappa_{0})^{\frac{1}{2}-\nu}-\sgn(g)(-ik/\kappa_{0})^{\frac{1}{2}+\nu}},\label{eq:32}
\end{align}
or, equivalently,
\begin{align}
  S_{\nu}(k)=
    \begin{dcases}
      -\frac{\sqrt{+ik/\kappa_{0}}\sin(i\nu\log(+ik/\kappa_{0}))}{\sqrt{-ik/\kappa_{0}}\sin(i\nu\log(-ik/\kappa_{0}))}&\text{for $g>0$},\\
      -\frac{\sqrt{+ik/\kappa_{0}}\cos(i\nu\log(+ik/\kappa_{0}))}{\sqrt{-ik/\kappa_{0}}\cos(i\nu\log(-ik/\kappa_{0}))}&\text{for $g<0$}.\\
    \end{dcases}\label{eq:33}
\end{align}
This is the S-matrix for the inverse-square-potential problem in the
intermediate window. As easily checked, this S-matrix satisfies the
following properties:\footnote{The Hermitian analyticity
  \eqref{eq:34b} is again just the Schwarz reflection principle. To
  see this, consider the scattering solution $\psi_{k}(r)$ for complex
  $k$, which is a linear combination of two independent solutions
  $f_{\nu}(kr)$ and $\overline{f_{\nu}(kr)}$ and given by
  $\psi_{k}(r)=\overline{f_{\nu}(k)}+S_{\nu}(k)f_{\nu}(k)$. In this
  case, the same calculation as \eqref{eq:30} and \eqref{eq:31} gives
  $S_{\nu}(k)=-g(-\overline{k})/g(k)$, where
  $g(k)=(\e^{-i\pi/2}k/\kappa_{0})^{\frac{1}{2}-\nu}-\sgn(g)(\e^{-i\pi/2}k/\kappa_{0})^{\frac{1}{2}+\nu}$. Note
  that this multivalued function can be real for $\arg k=\pi/2$ such
  that it satisfies the Schwarz reflection principle
  $g(k)=\overline{g(-\overline{k})}$ on the Riemann sheet containing
  $\arg k=\pi/2$. By using this, the S-matrix can be written as
  $S_{\nu}(k)=-g(-\overline{k})/g(k)=-\overline{g(k)}/g(k)$, which
  satisfies $S_{\nu}(k)=\overline{S_{\nu}(-\overline{k})}$ on such a
  Riemann sheet.}
\begin{subequations}
  \begin{itemize}
  \item\textbf{Property 1 (Unitarity).}
    \begin{align}
      \overline{S_{\nu}(k)}S_{\nu}(k)=1.\label{eq:34a}
    \end{align}
  \item\textbf{Property 2 (Hermitian analyticity).}
    \begin{align}
      \overline{S_{\nu}(k)}=S_{\nu}(-k).\label{eq:34b}
    \end{align}
  \item\textbf{Property 3 (Discrete phase invariance).}
    \begin{align}
      S_{\nu}(\e^{i\frac{n\pi}{\nu}}k)=S_{\nu}(k).\label{eq:34c}
    \end{align}
  \item\textbf{Property 4 (Circularly distributed simple poles).}
    \begin{alignat}{3}
      \lim_{k\to i\kappa_{0}\e^{in\pi/\nu}}(k-i\kappa_{0}\e^{i\frac{n\pi}{\nu}})S_{\nu}(k)&=i|N_{\kappa_{0}}|^{2}\e^{i\frac{n\pi}{\nu}}&&\quad\text{for}\quad g>0,&\label{eq:34d}\\
      \lim_{k\to i\kappa_{0}\e^{i(n+1/2)\pi/\nu}}(k-i\kappa_{0}\e^{i\frac{(n+1/2)\pi}{\nu}})S_{\nu}(k)&=i|N_{\kappa_{0}}|^{2}\e^{i\frac{(n+1/2)\pi}{\nu}}&&\quad\text{for}\quad g<0.&\label{eq:34e}
    \end{alignat}
  \end{itemize}
\end{subequations}
Here $k\in\mathbb{R}$ in \eqref{eq:34a}--\eqref{eq:34c} and
$n\in\mathbb{Z}$. We note that $S_{\nu}(k)$ is a multivalued function
of $k$ and has branch points at $k=0$ and $\infty$. Its Riemann
surface is infinitely sheeted if $\nu$ is irrational and is finitely
sheeted if $\nu$ is rational. A simple pole on the positive imaginary
axis on the first Riemann sheet is only the bound-state pole at
$k=i\kappa_{0}$ for $g>0$; see figure \ref{figure:2}.

Finally, we note that the above energy eigenfunctions provide a
complete orthonormal system of the problem. For instance, for $g>0$,
$\psi_{\kappa_{0}}$ and $\{\psi_{k}\}_{k>0}$ satisfy the
orthonormality conditions
\begin{subequations}
  \begin{align}
    &\int_{0}^{\infty}\!\!dr\,|\psi_{\kappa_{0}}(r)|^{2}=1,\label{eq:35a}\\
    &\int_{0}^{\infty}\!\!dr\,\overline{\psi_{k}(r)}\psi_{k^{\prime}}(r)=2\pi\delta(k-k^{\prime}),\quad\forall k,k^{\prime}>0,\label{eq:35b}\\
    &\int_{0}^{\infty}\!\!dr\,\overline{\psi_{\kappa_{0}}(r)}\psi_{k}(r)=\int_{0}^{\infty}\!\!dr\,\overline{\psi_{k}(r)}\psi_{\kappa_{0}}(r)=0,\quad\forall k>0,\label{eq:35c}
  \end{align}
\end{subequations}
as well as the completeness relation
\begin{align}
  \psi_{\kappa_{0}}(r)\overline{\psi_{\kappa_{0}}(r^{\prime})}+\int_{0}^{\infty}\!\!\frac{dk}{2\pi}\,\psi_{k}(r)\overline{\psi_{k}(r^{\prime})}=\delta(r-r^{\prime}),\quad\forall r,r^{\prime}>0.\label{eq:36}
\end{align}
Similarly, for $g<0$, $\{\psi_{k}\}_{k>0}$ satisfy the orthonormality
condition \eqref{eq:35b} as well as the completeness relation
$\int_{0}^{\infty}\!\!\frac{dk}{2\pi}\,\psi_{k}(r)\overline{\psi_{k}(r^{\prime})}=\delta(r-r^{\prime})$. For
the proof of these relations, see appendix \ref{appendix:A}. An
important point to note here is that all the complex energy solutions
corresponding to the simple poles on the higher Riemann sheets are
unphysical states and do not contribute to the completeness relation
\eqref{eq:36}.

\section{Few-body examples}
\label{section:3}
In many-body problems, the inverse-square potential naturally appears
as the centrifugal potential for the radial direction of many-body
configuration space. In this context, the coupling constant $\lambda$
in \eqref{eq:1} is basically the square of (hyper)angular momentum,
which must be quantized and cannot take arbitrary values. However,
just like the Aharonov-Bohm effect, if there exists a codimension-two
``magnetic'' flux in the configuration space, the (hyper)angular
momentum is shifted by the amount of flux and hence $\lambda$ can take
arbitrary values. In the following, we first revisit the one-body
Aharonov-Bohm problem in two dimensions
\cite{Giacconi:1995cs,Adami:1997ib} from the viewpoint of discrete
phase invariance. We then present a two-dimensional two-body problem
and a one-dimensional three-body problem that can exhibit scale
invariance breaking and discrete phase invariance.

\subsection{One-body Aharonov-Bohm problem}
\label{section:3.1}
Let us first consider a single charged particle of mass $m$ with
charge $q$ on the $xy$-plane in the presence of an infinitely thin
magnetic flux $\Phi$ at the origin. The Hamiltonian of this
two-dimensional system is given by
\begin{align}
  H_{\text{AB}}=-\frac{\hbar^{2}}{2m}\left(\bm{\nabla}+\frac{iq}{\hbar}\bm{A}\right)^{2},\label{eq:37}
\end{align}
where $\bm{\nabla}=(\partial/\partial x,\partial/\partial y)$. Here
$\bm{A}=(A_{x},A_{y})$ is a background gauge field satisfying
$\partial_{x}A_{y}-\partial_{y}A_{x}=\Phi\delta(x)\delta(y)$. In the
polar coordinate system $(r,\theta)$ given by
$(x,y)=(r\cos\theta,r\sin\theta)$, such a gauge field $\bm{A}$ can be
written as
\begin{align}
  \bm{A}=\frac{\Phi}{2\pi}\bm{\nabla}\theta.\label{eq:38}
\end{align}
In this gauge, the Hamiltonian \eqref{eq:37} can be cast into the
following form:
\begin{align}
  H_{\text{AB}}=\frac{\hbar^{2}}{2m}r^{-1/2}\left(-\frac{\partial^{2}}{\partial r^{2}}+\frac{(-i\partial_{\theta}+\alpha)^{2}-1/4}{r^{2}}\right)r^{1/2},\quad\text{where}\quad\alpha=\frac{q\Phi}{2\pi\hbar}.\label{eq:39}
\end{align}
Clearly, this Hamiltonian transforms as
$H_{\text{AB}}\mapsto \e^{-2t}H_{\text{AB}}$ under the scale
transformation $(r,\theta)\mapsto(\e^{t}r,\theta)$ so that the energy
eigenfunction $\psi_{E}(r,\theta)$ satisfying
$H_{\text{AB}}\psi_{E}(r,\theta)=E\psi_{E}(r,\theta)$ is expected to
fulfill the scaling law
$\psi_{\e^{2t}E}(r,\theta)\propto\psi_{E}(\e^{t}r,\theta)$ for
arbitrary $t$. This continuous scale invariance, however, can be
violated to discrete phase invariance if some eigenvalues of the
operator $(-i\partial_{\theta}+\alpha)^{2}$ are in the range
$(0,1)$. In the following, we discuss this scale invariance breaking
and its impact on scattering amplitude by analyzing the asymptotic
behavior of scattering solution. For simplicity, we will assume
rotational invariance (i.e., periodic boundary condition along the
$\theta$ direction).\footnote{Without rotational invariance, the
  Aharonov-Bohm Hamiltonian \eqref{eq:37} is known to possess a $U(2)$
  family of self-adjoint extensions \cite{Adami:1997ib}. In this
  paper, we will not discuss this generalization.}

First, under the periodic boundary condition, the energy eigenfunction
$\psi_{k}(r,\theta)$ with energy eigenvalue $E=k^{2}$ ($k>0$) can be
expanded into the following Fourier series:
\begin{align}
  \psi_{k}(r,\theta)=\frac{1}{\sqrt{kr}}\sum_{n=-\infty}^{\infty}C_{n}\psi_{k,n}(r)\frac{\e^{in\theta}}{\sqrt{2\pi}},\label{eq:40}
\end{align}
where $\{C_{n}\}_{n=-\infty}^{\infty}$ are expansion coefficients. The
factor $1/\sqrt{kr}$ is introduced just for later convenience. Then,
the energy eigenvalue equation $H_{\text{AB}}\psi_{k}=k^{2}\psi_{k}$
is reduced to the following:
\begin{align}
  \left(-\frac{d^{2}}{dr^{2}}+\frac{(n+\alpha)^{2}-1/4}{r^{2}}\right)\psi_{k,n}(r)=k^{2}\psi_{k,n}(r).\label{eq:41}
\end{align}
It is now clear that the problem is equivalent to the
inverse-square-potential problem with $\nu=|n+\alpha|$ in the previous
section. Hence, for $0<|n+\alpha|<1$ (i.e., for $n=-[\alpha]$ and
$n=-[\alpha]-1$, where $[\alpha]$ stands for the greatest integer
lower than or equal to $\alpha$), $\psi_{k,n}$ is given by
\eqref{eq:29}. For $|n+\alpha|>1$, on the other hand, the scattering
solution is given in terms of the Bessel function of the first kind by
$\psi_{k,n}(r)=\e^{-i(2|n+\alpha|+1)/4}\sqrt{2\pi
  kr}J_{|n+\alpha|}(kr)$, which behaves as
$\psi_{k,n}(r)\to\e^{-ikr}-i\e^{-i|n+\alpha|\pi}\e^{ikr}$ as
$r\to\infty$. The energy eigenfunction \eqref{eq:40} then takes the
following asymptotic form:
\begin{align}
  \psi_{k}(r,\theta)\to\frac{1}{\sqrt{kr}}\sum_{n=-\infty}^{\infty}C_{n}\left(\e^{-ikr}+S_{|n+\alpha|}(k)\e^{ikr}\right)\frac{\e^{in\theta}}{\sqrt{2\pi}}\quad\text{as}\quad r\to\infty,\label{eq:42}
\end{align}
where
\begin{align}
  S_{|n+\alpha|}(k)=
  \begin{dcases}
    -\frac{(+ik/\kappa_{n})^{\frac{1}{2}-|n+\alpha|}-\sgn(g)(+ik/\kappa_{n})^{\frac{1}{2}+|n+\alpha|}}{(-ik/\kappa_{n})^{\frac{1}{2}-|n+\alpha|}-\sgn(g)(-ik/\kappa_{n})^{\frac{1}{2}+|n+\alpha|}}&\text{for $n=-[\alpha]$ and $n=-[\alpha]-1$},\\
    -i\e^{-i|n+\alpha|\pi}&\text{otherwise}.\\
  \end{dcases}\label{eq:43}
\end{align}
Here $\kappa_{n}$ ($n=-[\alpha],-[\alpha]-1$) are two independent
positive parameters.

On the other hand, the Green's function analysis tells us that the
energy eigenfunction takes the following asymptotic form:
\begin{align}
  \psi_{k}(r,\theta)\to\e^{ikr\cos\theta}+f(k,\theta)\frac{\e^{i(kr+\pi/4)}}{r^{1/2}}\quad\text{as}\quad r\to\infty,\label{eq:44}
\end{align}
where $f(k,\theta)$ is the scattering amplitude whose modulus squared
gives the differential cross section:
\begin{align}
  \frac{d\sigma}{d\theta}(k,\theta)=|f(k,\theta)|^{2}.\label{eq:45}
\end{align}
This scattering amplitude can be obtained by comparing \eqref{eq:44}
with \eqref{eq:42}. By using the Fourier series,
\begin{subequations}
  \begin{align}
    \e^{ikr\cos\theta}&=\sum_{n=-\infty}^{\infty}i^{|n|}J_{|n|}(kr)\e^{in\theta},\label{eq:46a}\\
    f(k,\theta)&=\sum_{n=-\infty}^{\infty}f_{n}(k)\frac{\e^{in\theta}}{\sqrt{2\pi}},\label{eq:46b}
  \end{align}
\end{subequations}
as well as the asymptotic behavior of the Bessel function,
\begin{align}
  J_{\nu}(z)\to\frac{1}{\sqrt{2\pi z}}\left(\e^{-i(z-\frac{2\nu+1}{4}\pi)}+\e^{i(z-\frac{2\nu+1}{4}\pi)}\right)\quad\text{as}\quad|z|\to\infty,\label{eq:47}
\end{align}
we find that \eqref{eq:44} can be written as the following Fourier
series:
\begin{align}
  \psi_{k}(r,\theta)\to\frac{1}{\sqrt{kr}}\sum_{n=-\infty}^{\infty}i^{|n|}\left[\e^{-i(kr-\frac{2|n|+1}{4}\pi)}+\left(1+i\sqrt{k}f_{n}(k)\right)\e^{i(kr-\frac{2|n|+1}{4}\pi)}\right]\frac{\e^{in\theta}}{\sqrt{2\pi}}\quad\text{as}\quad r\to\infty.\label{eq:48}
\end{align}
Comparing \eqref{eq:42} with \eqref{eq:48}, we immediately find
$C_{n}=i^{|n|}\e^{i((2|n|+1)/4)\pi}=(-1)^{|n|}\e^{i\pi/4}$ and
$S_{|n+\alpha|}(k)=(1+i\sqrt{k}f_{n}(k))\e^{-i(|n|+1/2)\pi}$, from
which we get the following partial-wave amplitude:
\begin{align}
  f_{n}(k)=\frac{S_{|n+\alpha|}(k)\e^{i(|n|+\frac{1}{2})\pi}-1}{i\sqrt{k}}.\label{eq:49}
\end{align}
Note that these partial-wave amplitudes are consistent with the known
result \cite{Giacconi:1995cs,Adami:1997ib} when combined with
\eqref{eq:46b}.

Now, it is clear from \eqref{eq:43} that \eqref{eq:49} satisfies the
following complex scaling law:
\begin{align}
  f_{n}(\e^{i\frac{\ell\pi}{|n+\alpha|}}k)=\e^{-i\frac{\ell\pi}{2|n+\alpha|}}f_{n}(k)\quad\text{for}\quad n=-[\alpha],-[\alpha]-1,\label{eq:50}
\end{align}
where $\ell$ is an arbitrary integer. In addition, it possesses
infinitely many simple poles in the complex $k$-plane and satisfies
\begin{align}
  f_{n}(k)\to\frac{1}{\sqrt{\kappa_{n}}}\frac{|N_{\kappa_{n}}|^{2}\e^{i\frac{\ell\pi}{2|n+\alpha|}}}{k-i\kappa_{n}\e^{i\frac{\ell\pi}{|n+\alpha|}}}+O(1)\label{eq:51}
\end{align}
as
\begin{align}
  k\to i\kappa_{n}\e^{i\frac{\ell\pi}{|n+\alpha|}}=-\kappa_{n}\sin\left(\frac{\ell\pi}{|n+\alpha|}\right)+i\kappa_{n}\cos\left(\frac{\ell\pi}{|n+\alpha|}\right)\label{eq:52}
\end{align}
for $n=-[\alpha],-[\alpha]-1$. Notice that the above simple pole is
located near the positive real $k$-axis if
$\sin(\ell\pi/|n+\alpha|)<0$ and $|\cos(\ell\pi/|n+\alpha|)|\ll1$. In
such a case, the scattering amplitude may well experience a
resonance-like enhancement when the positive real wavenumber $k$
passes through the point $k=-\kappa_{n}\sin(\ell\pi/|n+\alpha|)>0$.

To summarize, continuous scale invariance in the (rotationally
invariant) Aharonov-Bohm problem can be broken to discrete phase
invariance in two channels $n=-[\alpha]$ and $-[\alpha]-1$ thanks to
the interplay between the magnetic flux at $r=0$ and the boundary
condition at $r=0$. Though the Aharonov-Bohm scattering with
nontrivial boundary condition has long been studied in the literature
(especially in the context of self-adjoint extension)
\cite{Giacconi:1995cs,Adami:1997ib}, its invariance under the discrete
phase transformation and the appearance of circularly distributed
simple poles have not been appreciated before.

The above one-body Aharonov-Bohm problem can be easily embedded into
many-body problems if there is a codimension-two ``magnetic'' flux in
the many-body configuration space. Typical examples are a
two-dimensional two-body problem and a one-dimensional three-body
problem of nonidentical particles, where the ``magnetic'' fluxes are
penetrating through the two-body and three-body coincidence points,
respectively. In the subsequent subsections, we will study these
few-body problems and point out that continuous scale invariance can
be broken to discrete phase invariance in exactly the same way as for
the Aharonov-Bohm problem.

\subsection{Two-body problem in two dimensions}
\label{section:3.2}
In general, one-body problem can be regarded as relative motion of
two-body problem. From this viewpoint, it is almost trivial that
scale invariance breaking in the one-body Aharonov-Bohm problem can be
applied to the two-body problem as well. In this subsection, however,
we will revisit this problem from the viewpoint of topology of
configuration space. As we will see in the next subsection, the
topological analysis below is easily generalized to the three-body
problem in one dimension.

To begin with, let us consider two nonidentical particles of masses
$m_{1}$ and $m_{2}$, each of which are located at positions
$\bm{r}_{1}$ and $\bm{r}_{2}$ in the two-dimensional plane
$\mathbb{R}^{2}$. Suppose that the two-body wavefunction becomes
singular at the two-body coincidence point $\bm{r}_{1}=\bm{r}_{2}$. In
this case, the two-body configuration space is given by removing the
singular points and defined by the following subtracted space:
\begin{align}
  M_{\text{2-body}}\coloneq\mathbb{R}^{2}\times\mathbb{R}^{2}-\Delta_{2},\label{eq:53}
\end{align}
where $\Delta_{2}$ stands for the set of two-body coincidence points
defined by
\begin{align}
  \Delta_{2}\coloneq\{(\bm{r}_{1},\bm{r}_{2})\in\mathbb{R}^{2}\times\mathbb{R}^{2}:\bm{r}_{1}=\bm{r}_{2}\}.\label{eq:54}
\end{align}
As we will see soon, $\Delta_{2}$ turns out to describe the support of
infinitely-thin magnetic flux penetrating through the two-body
coincidence points. For two-body wavefunctions, $\Delta_{2}$ turns out
to become the branch-point singularities. Hence, eq.~\eqref{eq:53}
physically describes the configuration space of two nonidentical
anyons. For more details of nonidentical anyons, see, e.g.,
\cite{Jo:1996}.

Now, as is well known, the above configuration space is multiply
connected. To see this, let us introduce the relative coordinate
$\bm{\xi}_{1}$ and the center-of-mass coordinate $\bm{\xi}_{2}$ by
\begin{align}
  \bm{\xi}_{1}\coloneq\bm{r}_{1}-\bm{r}_{2},\quad
  \bm{\xi}_{2}\coloneq\frac{m_{1}\bm{r}_{1}+m_{2}\bm{r}_{2}}{m_{1}+m_{2}}.\label{eq:55}
\end{align}
Note that the two-body coincidence point $\bm{r}_{1}=\bm{r}_{2}$
corresponds to $\bm{\xi}_{1}=\bm{0}$. Hence, in the configuration
space, $\bm{\xi}_{1}=\bm{\bm{0}}$ is excluded while $\bm{\xi}_{2}$ can
take arbitrary value. The two-body configuration space \eqref{eq:53}
is thus factorized as follows:
\begin{align}
  M_{\text{2-body}}\cong\mathring{\mathbb{R}}^{2}\times\mathbb{R}^{2},\label{eq:56}
\end{align}
where the first factor
$\mathring{\mathbb{R}}^{2}\coloneq\{\bm{\xi}_{1}\in\mathbb{R}^{2}:\bm{\xi}_{1}\neq\bm{0}\}$
describes the space of relative motion and the second factor
$\mathbb{R}^{2}\ni\bm{\xi}_{2}$ describes the space of center-of-mass
motion. Note that $\mathring{\mathbb{R}}^{2}$ is the one-punctured
plane so that its fundamental group is the additive group of integer
$\mathbb{Z}$. Thus we find
\begin{align}
  \pi_{1}(M_{\text{2-body}})\cong\pi_{1}(\mathring{\mathbb{R}}^{2})\oplus\pi_{1}(\mathbb{R}^{2})\cong\mathbb{Z}\oplus\{1\}\cong\mathbb{Z}.\label{eq:57}
\end{align}
Physically, this group describes the winding number of two-particle
trajectories around the two-body coincidence point. Theoretically,
this group appears as boundary condition for two-body
wavefunctions. Indeed, according to the Dowker's covering-space method
\cite{Dowker:1972np}, the two-body wavefunction satisfies
$\Psi(g^{n}\bm{r}_{1},g^{n}\bm{r}_{2})=D(g^{n})\Psi(\bm{r}_{1},\bm{r}_{2})$,
where $g$ is the generator of $\mathbb{Z}$,
$(g^{n}\bm{r}_{1},g^{n}\bm{r}_{2})$ stands for the action of the group
$\mathbb{Z}$ on the coordinates $(\bm{r}_{1},\bm{r}_{2})$ with $n$
being an integer, and $D$ is a one-dimensional unitary representation
of $\mathbb{Z}$. We note that, for the additive group $\mathbb{Z}$,
there is a one-parameter family of one-dimensional unitary
representations given by $D(g^{n})=\e^{i2\pi n\alpha}$, where $\alpha$
is a real parameter.

To find out the boundary condition more explicitly, suppose that the
system is invariant under the translation. In such a
translation-invariant system, the two-body wavefunction is factorized
as $\Psi(\bm{r}_{1},\bm{r}_{2})=\psi(\bm{\xi}_{1})\phi(\bm{\xi}_{2})$,
where $\psi$ and $\phi$ are the wavefunctions of relative and
center-of-mass motions, respectively. Let us next introduce the polar
coordinate system $(r,\theta)$ in $\mathring{\mathbb{R}}^{2}$ by
\begin{align}
  \bm{\xi}_{1}=(r\cos\theta,r\sin\theta).\label{eq:58}
\end{align}
In this polar coordinate system, the action of the group $\mathbb{Z}$
on the coordinates is given by
$(r,\theta,\bm{\xi}_{2})\to(r,\theta+2n\pi,\bm{\xi}_{2})$. Hence, the
center-of-mass wavefunction $\phi(\bm{\xi}_{2})$ remains unchanged
while the relative wavefunction $\psi(r,\theta)$ undergoes the
following twisted boundary condition:
\begin{align}
  \psi(r,\theta+2n\pi)=\e^{i2n\pi\alpha}\psi(r,\theta).\label{eq:59}
\end{align}
It is now clear that $\psi$ is a multivalued function of $\theta$ with
the branch-point singularity at $r=0$. Physically, the boundary
condition \eqref{eq:59} describes the Aharonov-Bohm phase for a
two-particle trajectory winding around the ``magnetic'' flux $n$
times, where the flux is penetrating through $r=0$ and proportional to
$\alpha$. As in the case of identical anyons, this Aharonov-Bohm phase
could be realized by the Wilczek's charge-flux picture
\cite{Wilczek:1981du,Wilczek:1982wy}.

Now let us apply the above topological analysis to the
scale invariance breaking in two-body problems. As the simplest
example of scale-invariant two-body Hamiltonian, let us take the
following free Hamiltonian:
\begin{align}
  H_{\text{2-body}}=-\frac{\hbar^{2}}{2m_{1}}\bm{\nabla}_{\bm{r}_{1}}^{2}-\frac{\hbar^{2}}{2m_{2}}\bm{\nabla}_{\bm{r}_{2}}^{2},\label{eq:60}
\end{align}
where $\bm{\nabla}_{\bm{r_{j}}}^{2}$ stands for the Laplacian with
respect to the coordinates $\bm{r}_{j}$. By using the relative and
center-of-mass coordinates, this Hamiltonian is decomposed as
$H_{\text{2-body}}=H_{\text{rel}}+H_{\text{cm}}$. Here
$H_{\text{rel}}=-(\hbar^{2}/(2\mu_{1}))\bm{\nabla}_{\bm{\xi}_{1}}^{2}$
is the relative Hamiltonian and
$H_{\text{cm}}=-(\hbar^{2}/(2\mu_{2}))\bm{\nabla}_{\bm{\xi}_{2}}^{2}$
is the center-of-mass Hamiltonian, where
$\mu_{1}\coloneq(1/m_{1}+1/m_{2})^{-1}$ is the reduced mass and
$\mu_{2}\coloneq m_{1}+m_{2}$ is the total mass. In terms of the polar
coordinates, the relative Hamiltonian can be written as
\begin{align}
  H_{\text{rel}}=\frac{\hbar^{2}}{2\mu_{1}}r^{-1/2}\left(-\frac{\partial^{2}}{\partial r^{2}}+\frac{-\partial_{\theta}^{2}}{r^{2}}\right)r^{1/2}.\label{eq:61}
\end{align}
Let $\psi_{k}$ be the relative wavefunction satisfying the eigenvalue
equation $H_{\text{rel}}\psi_{k}=k^{2}\psi_{k}$. Then, under the
twisted boundary condition \eqref{eq:59}, the relative wavefunction is
decomposed as
\begin{align}
  \psi_{k}(r,\theta)=\frac{1}{\sqrt{kr}}\sum_{n=-\infty}^{\infty}C_{n}\psi_{k,n}(r)\frac{\e^{i(n+\alpha)\theta}}{\sqrt{2\pi}}.\label{eq:62}
\end{align}
Substituting this into the eigenvalue equation
$H_{\text{rel}}\psi_{k}=k^{2}\psi_{k}$, we get the same radial
Schr\"{o}dinger equation as \eqref{eq:41}, meaning that the problem is
equivalent to the one-body Aharonov-Bohm problem in the previous
subsection. Hence the continuous scale invariance can be broken to
discrete phase invariance in exactly the same way as for the
Aharonov-Bohm problem.

In the next subsection, we will show that the above analysis can be
applied to one-dimensional three-body problems as well.

\subsection{Three-body problem in one dimension}
\label{section:3.3}
Let us consider three nonidentical particles of masses $m_{1}$,
$m_{2}$, and $m_{3}$, each of which are located at $x_{1}$, $x_{2}$,
and $x_{3}$ in the one-dimensional line $\mathbb{R}$. Suppose that the
three-body wavefunction is singular at the three-body coincidence
point $x_{1}=x_{2}=x_{3}$. In this case, the three-body configuration
space is given by the following subtracted space:
\begin{align}
  M_{\text{3-body}}\coloneq\mathbb{R}\times\mathbb{R}\times\mathbb{R}-\Delta_{3},\label{eq:63}
\end{align}
where $\Delta_{3}$ stands for the set of three-body coincidence points
defined by
\begin{align}
  \Delta_{3}\coloneq\{(x_{1},x_{2},x_{3})\in\mathbb{R}\times\mathbb{R}\times\mathbb{R}:x_{1}=x_{2}=x_{3}\}.\label{eq:64}
\end{align}
This configuration space is again multiply connected. To see this, let
us introduce the relative and center-of-mass coordinates as follows:
\begin{align}
  \xi_{1}\coloneq x_{1}-x_{2},\quad
  \xi_{2}\coloneq\frac{m_{1}x_{1}+m_{2}x_{2}}{m_{1}+m_{2}}-x_{3},\quad
  \xi_{3}\coloneq\frac{m_{1}x_{1}+m_{2}x_{2}+m_{3}x_{3}}{m_{1}+m_{2}+m_{3}}.\label{eq:65}
\end{align}
Note that the condition $x_{1}=x_{2}=x_{3}$ is equivalent to the
condition $(\xi_{1},\xi_{2})=(0,0)$, from which we find that the
configuration space is factorized as follows:
\begin{align}
  M_{\text{3-body}}\cong\mathring{\mathbb{R}}^{2}\times\mathbb{R},\label{eq:66}
\end{align}
where
$\mathring{\mathbb{R}}^{2}=\{(\xi_{1},\xi_{2})\in\mathbb{R}^{2}:(\xi_{1},\xi_{2})\neq(0,0)\}$
is the space of relative motion and $\mathbb{R}\ni\xi_{3}$ is the
space of center-of-mass motion. Obviously, the fundamental group of
\eqref{eq:66} is the same as \eqref{eq:57} and given by
\begin{align}
  \pi_{1}(M_{\text{3-body}})\cong\mathbb{Z}.\label{eq:67}
\end{align}
Hence, the three-body wavefunction satisfies the same boundary
condition as in the previous subsection. Physically, eq.~\eqref{eq:63}
describes the three-body configuration space for the nonidentical
version of the so-called traid anyons \cite{Harshman:2018wzv}. For
more details of nonidentical traid anyons, we refer to
\cite{Ohya:2023tkd}.

To see the scale invariance breaking, let us consider the three-body
free Hamiltonian,
\begin{align}
  H_{\text{3-body}}=-\frac{\hbar^{2}}{2m_{1}}\frac{\partial^{2}}{\partial x_{1}^{2}}-\frac{\hbar^{2}}{2m_{2}}\frac{\partial^{2}}{\partial x_{2}^{2}}-\frac{\hbar^{2}}{2m_{3}}\frac{\partial^{2}}{\partial x_{3}^{2}}.\label{eq:68}
\end{align}
In terms of the coordinate system $(\xi_{1},\xi_{2},\xi_{3})$, this
Hamiltonian is decomposed as
$H_{\text{3-body}}=H_{\text{rel}}+H_{\text{cm}}$, where
$H_{\text{rel}}\coloneq-(\hbar^{2}/(2\mu_{1}))\partial^{2}/\partial\xi_{1}^{2}-(\hbar^{2}/(2\mu_{2}))\partial^{2}/\partial\xi_{2}^{2}$
and
$H_{\text{cm}}\coloneq-(\hbar^{2}/(2\mu_{3}))\partial^{2}/\partial\xi_{3}^{2}$
with $\mu_{1}\coloneq(1/m_{1}+1/m_{2})^{-1}$,
$\mu_{2}\coloneq(1/(m_{1}+m_{2})+1/m_{3})^{-1}$, and
$\mu_{3}\coloneq m_{1}+m_{2}+m_{3}$. Let us introduce the polar
coordinate system $(r,\theta)$ in $\mathring{\mathbb{R}}^{2}$ by
\begin{align}
  (\xi_{1},\xi_{2})=\left(\sqrt{\frac{\mu_{0}}{\mu_{1}}}r\cos\theta,\sqrt{\frac{\mu_{0}}{\mu_{2}}}r\sin\theta\right),\label{eq:69}
\end{align}
where $\mu_{0}(>0)$ is an arbitrary reference mass scale. In terms of
this polar coordinate system, the relative Hamiltonian takes the
following form:
\begin{align}
  H_{\text{rel}}=\frac{\hbar^{2}}{2\mu_{0}}r^{-1/2}\left(-\frac{\partial^{2}}{\partial r^{2}}+\frac{-\partial_{\theta}^{2}}{r^{2}}\right)r^{1/2}.\label{eq:70}
\end{align}
It is now clear that continuous scale invariance can be broken in the
one-dimensional three-body problem as well. Indeed, by decomposing the
three-body wavefunction into the product of relative and
center-of-mass wavefunctions,
$\Psi(x_{1},x_{2},x_{3})=\psi(\xi_{1},\xi_{2})\phi(\xi_{3})$, and
writing the relative wavefunction in terms of the polar coordinates
$(r,\theta)$, we get the same twisted boundary condition as well as
the same radial Schr\"{o}dinger equation as in the previous
subsection. Hence, continuous scale invariance can be broken to
discrete phase invariance in exactly the same way as for the
Aharonov-Bohm problem.

\section{Conclusion}
\label{section:4}
Continuous scale invariance can be broken to some discrete symmetry in
quantum mechanics. A well-known example is the breakdown to discrete
scale invariance, which has been widely studied over the years in
various contexts of physics. It was, however, not appreciated in the
literature that there is another discrete symmetry to which continuous
scale invariance can be broken. That is the discrete phase invariance,
or discrete $U(1)$ symmetry, which can be realized as a
complexification of discrete scale invariance.

In this paper, we have discussed several few-body examples that
exhibit the breakdown of continuous scale invariance to discrete phase
invariance. We have first revisited the one-body
inverse-square-potential problem as a prototypical example of
scale invariance breaking in quantum mechanics. We have shown that,
when the coupling constant is in the intermediate window between the
lower and upper critical values, continuous scale invariance can be
broken to discrete phase invariance. In this case, discrete phase
invariance manifests itself as circularly distributed simple poles on
the Riemann surface of the S-matrix, where the S-matrix is a
multivalued function of momentum. We have then applied this result to
the two-dimensional two-body problem and one-dimensional three-body
problem of nonidentical particles with codimension-two flux, both of
which can be reduced to the one-body Aharonov-Bohm problem. In these
example, we have shown that discrete phase invariance can be realized
in two particular channels of partial wave.

We note that the Aharonov-Bohm problem and its variants are merely
examples that can be reduced to the inverse-square-potential problem
in the intermediate window. It would be interesting to study other
models as well. For instance, it would be worth investigating the
possibility of realizing discrete phase invariance in the three-body
problem of two heavy and one light fermions discussed in
\cite{Nishida:2007mr}. Also, it would be important to investigate
physical implications of simple poles on the higher Riemann sheets
more closely.

\subsection*{Acknowledgment}
This work was supported by JSPS KAKENHI Grant Number JP23K03267.

\begin{appendices}
  \setcounter{equation}{0}
  \renewcommand{\theequation}{\thesection.\arabic{equation}}
  \section{Proof of the orthonormality and completeness}
  \label{appendix:A}
  In this section, we prove the orthonormality and completeness of the
  energy eigenfunctions for the inverse-square-potential problem in
  the intermediate window with the boundary condition
  \eqref{eq:20}. For brevity, we just focus on the case $g>0$. The
  case $g<0$ can be proved in a similar manner.

  \subsection{Orthonormality}
  \label{appendix:A.1}
  We first prove the orthonormality conditions
  \eqref{eq:35a}--\eqref{eq:35c}. The keys for the proof are the
  asymptotic behaviors of energy eigenfunctions at infinity and the
  boundary condition at $r=0$. To see this, let $\psi_{E}$ and
  $\psi_{E^{\prime}}$ be two distinct energy eigenfunctions satisfying
  $H\psi_{E}=E\psi_{E}$ and
  $H\psi_{E^{\prime}}=E^{\prime}\psi_{E^{\prime}}$, where $H$ is given
  by \eqref{eq:1} and $E$ and $E^{\prime}$ are its two distinct real
  eigenvalues. Then the following identity holds:
  \begin{align}
    \int_{0}^{\infty}\!\!dr\,\overline{\psi_{E}(r)}\psi_{E^{\prime}}(r)
    =\frac{1}{E^{\prime}-E}\int_{0}^{\infty}\!\!dr\left[\overline{\psi_{E}(r)}H\psi_{E^{\prime}}(r)-\left(H\overline{\psi_{E}(r)}\right)\psi_{E^{\prime}}(r)\right].\label{eq:A.1}
  \end{align}
  Notice that the integrand on the right-hand side can be written as a
  total derivative. In fact, by using the factorization \eqref{eq:13},
  we have
  \begin{align}
    \overline{\psi_{E}(r)}H\psi_{E^{\prime}}(r)-\left(H\overline{\psi_{E}(r)}\right)\psi_{E^{\prime}}(r)
    &=\frac{d}{dr}\left[-\frac{\overline{\psi_{E}(r)}}{r^{\frac{1}{2}-\nu}}\left(r^{1-2\nu}\frac{d}{dr}\frac{\psi_{E^{\prime}}(r)}{r^{\frac{1}{2}-\nu}}\right)
      +\left(r^{1-2\nu}\frac{d}{dr}\frac{\overline{\psi_{E}(r)}}{r^{\frac{1}{2}-\nu}}\right)\frac{\psi_{E^{\prime}}(r)}{r^{\frac{1}{2}-\nu}}\right]\nonumber\\
    &=\frac{d}{dr}\left[-\overline{\psi_{E}(r)}\frac{d\psi_{E^{\prime}}(r)}{dr}+\frac{d\overline{\psi_{E}(r)}}{dr}\psi_{E^{\prime}}(r)\right].\label{eq:A.2}
  \end{align}
  Putting this into \eqref{eq:A.1}, we get
  \begin{align}
    \int_{0}^{\infty}\!\!dr\,\overline{\psi_{E}(r)}\psi_{E^{\prime}}(r)
    &=\frac{1}{E^{\prime}-E}\lim_{r\to0}\left[\frac{\overline{\psi_{E}(r)}}{r^{\frac{1}{2}-\nu}}\left(r^{1-2\nu}\frac{d}{dr}\frac{\psi_{E^{\prime}}(r)}{r^{\frac{1}{2}-\nu}}\right)
      -\left(r^{1-2\nu}\frac{d}{dr}\frac{\overline{\psi_{E}(r)}}{r^{\frac{1}{2}-\nu}}\right)\frac{\psi_{E^{\prime}}(r)}{r^{\frac{1}{2}-\nu}}\right]\nonumber\\
    &\quad+\frac{1}{E^{\prime}-E}\lim_{r\to\infty}\left[-\overline{\psi_{E}(r)}\frac{d\psi_{E^{\prime}}(r)}{dr}+\frac{d\overline{\psi_{E}(r)}}{dr}\psi_{E^{\prime}}(r)\right],\label{eq:A.3}
  \end{align}
  where we have used the first line of \eqref{eq:A.2} for the surface
  term at $r=0$ and the second line of \eqref{eq:A.2} for the surface
  term at $r=\infty$. Now it is clear that the orthonormality
  conditions can be proved by just evaluating the boundary and
  asymptotic behaviors of energy eigenfunctions.

  Let us first prove the normalization condition \eqref{eq:35a} for
  the bound-state solution $\psi_{\kappa_{0}}$ by calculating the
  normalization factor. Let $\psi_{\kappa}$ be a negative-energy
  solution given by \eqref{eq:24}, where $\kappa$ is an arbitrary
  positive real here. Then, the squared norm of $\psi_{\kappa}$ is
  given by
  \begin{align}
    \int_{0}^{\infty}\!\!dr\,|\psi_{\kappa}(r)|^{2}
    &=\lim_{\kappa^{\prime}\to\kappa}\int_{0}^{\infty}\!\!dr\,\overline{\psi_{\kappa}(r)}\psi_{\kappa^{\prime}}(r)\nonumber\\
    &=\lim_{\kappa^{\prime}\to\kappa}\frac{1}{-\kappa^{\prime2}+\kappa^{2}}\lim_{r\to0}\left[\frac{\overline{\psi_{\kappa}(r)}}{r^{\frac{1}{2}-\nu}}\left(r^{1-2\nu}\frac{d}{dr}\frac{\psi_{\kappa^{\prime}}(r)}{r^{\frac{1}{2}-\nu}}\right)
      -\left(r^{1-2\nu}\frac{d}{dr}\frac{\overline{\psi_{\kappa}(r)}}{r^{\frac{1}{2}-\nu}}\right)\frac{\psi_{\kappa^{\prime}}(r)}{r^{\frac{1}{2}-\nu}}\right]\nonumber\\
    &=\lim_{\kappa^{\prime}\to\kappa}\frac{1}{-\kappa^{\prime2}+\kappa^{2}}\left[\kappa^{\frac{1}{2}-\nu}A_{\nu}\left(-2\nu\kappa^{\prime\frac{1}{2}+\nu}B_{\nu}\right)-\left(-2\nu\kappa^{\frac{1}{2}+\nu}B_{\nu}\right)\kappa^{\prime\frac{1}{2}-\nu}A_{\nu}\right]\overline{N_{\kappa}}N_{\kappa^{\prime}}\nonumber\\
    &=\lim_{\kappa^{\prime}\to\kappa}\frac{2\nu A_{\nu}B_{\nu}\sqrt{\kappa\kappa^{\prime}}((\kappa^{\prime}/\kappa)^{\nu}-(\kappa/\kappa^{\prime})^{\nu})}{\kappa^{\prime2}-\kappa^{2}}\overline{N_{\kappa}}N_{\kappa^{\prime}}\nonumber\\
    &=\frac{2\nu^{2}A_{\nu}B_{\nu}}{\kappa}|N_{\kappa}|^{2},\label{eq:A.4}
  \end{align}
  where the second equality follows from \eqref{eq:A.3} and the fact
  that the surface term at $r=\infty$ vanishes thanks to the
  asymptotic behaviors
  $\psi_{\kappa}(r),\psi_{\kappa^{\prime}}(r)\to0$ as
  $r\to\infty$. The third equality follows from the boundary behavior
  \eqref{eq:25}. Requiring
  $\int_{0}^{\infty}\!\!dr\,|\psi_{\kappa}(r)|^{2}=1$, we find
  \begin{align}
    |N_{\kappa}|=\sqrt{\frac{\kappa}{2\nu^{2}A_{\nu}B_{\nu}}}=\sqrt{\frac{\kappa\sin(\nu\pi)}{\nu}},\label{eq:A.5}
  \end{align}
  where we have used \eqref{eq:23} and the Euler's reflection formula
  $\Gamma(1+\nu)\Gamma(1-\nu)=\nu\pi/\sin(\nu\pi)$. Since
  \eqref{eq:A.4} holds for arbitrary $\kappa>0$, the normalization
  condition $\int_{0}^{\infty}\!\!dr\,|\psi_{\kappa}(r)|^{2}=1$ also
  holds for $\kappa=\kappa_{0}$. Thus we obtain \eqref{eq:35a}.

  Let us next prove the orthogonality \eqref{eq:35c} between the
  bound-state solution $\psi_{\kappa_{0}}$ and the scattering solution
  $\psi_{k}$. Substituting $\psi_{\kappa_{0}}$ for $\psi_{E}$ and
  $\psi_{k}$ for $\psi_{E^{\prime}}$ in \eqref{eq:A.3}, we get
  \begin{align}
    \int_{0}^{\infty}\!\!dr\,\overline{\psi_{\kappa_{0}}(r)}\psi_{k}(r)
    &=\frac{1}{k^{2}+\kappa_{0}^{2}}\lim_{r\to0}\left[\frac{\overline{\psi_{\kappa_{0}}(r)}}{r^{\frac{1}{2}-\nu}}\left(r^{1-2\nu}\frac{d}{dr}\frac{\psi_{k}(r)}{r^{\frac{1}{2}-\nu}}\right)
      -\left(r^{1-2\nu}\frac{d}{dr}\frac{\overline{\psi_{\kappa_{0}}(r)}}{r^{\frac{1}{2}-\nu}}\right)\frac{\psi_{k}(r)}{r^{\frac{1}{2}-\nu}}\right]\nonumber\\
    &\quad+\frac{1}{k^{2}+\kappa_{0}^{2}}\lim_{r\to\infty}\left[-\overline{\psi_{\kappa_{0}}(r)}\frac{d\psi_{k}(r)}{dr}+\frac{d\overline{\psi_{\kappa_{0}}(r)}}{dr}\psi_{k}(r)\right]\nonumber\\
    &=\frac{1}{k^{2}+\kappa_{0}^{2}}\lim_{r\to0}\left[-\frac{1}{g}\frac{\overline{\psi_{\kappa_{0}}(r)}}{r^{\frac{1}{2}-\nu}}\frac{\psi_{k}(r)}{r^{\frac{1}{2}-\nu}}
      +\frac{1}{g}\frac{\overline{\psi_{\kappa_{0}}(r)}}{r^{\frac{1}{2}-\nu}}\frac{\psi_{k}(r)}{r^{\frac{1}{2}-\nu}}\right]\nonumber\\
    &=0,\label{eq:A.6}
  \end{align}
  where the second equality follows from the boundary condition
  \eqref{eq:20} and the asymptotic behaviors
  $\psi_{\kappa_{0}}(r),d\psi_{\kappa_{0}}(r)/dr\to0$ as
  $r\to\infty$. Thus $\psi_{\kappa_{0}}$ and $\psi_{k}$ are orthogonal
  for arbitrary $k>0$. By taking the complex conjugate, we also get
  $\int_{0}^{\infty}\!\!dr\,\overline{\psi_{k}(r)}\psi_{\kappa_{0}}(r)=0$.
  Thus we obtain \eqref{eq:35c}.
  
  Let us finally prove the orthonormality \eqref{eq:35b} of the
  scattering solutions. Substituting $\psi_{k}$ for $\psi_{E}$ and
  $\psi_{k^{\prime}}$ for $\psi_{E^{\prime}}$ in \eqref{eq:A.3}, we
  get
  \begin{align}
    \int_{0}^{\infty}\!\!dr\,\overline{\psi_{k}(r)}\psi_{k^{\prime}}(r)
    &=\frac{1}{k^{\prime2}-k^{2}}\lim_{r\to\infty}\left(-\overline{\psi_{k}(r)}\frac{d\psi_{k^{\prime}}(r)}{dr}+\overline{\frac{d\psi_{k}(r)}{dr}}\psi_{k}(r)\right)\nonumber\\
    &=\frac{1}{k^{\prime2}-k^{2}}\lim_{r\to\infty}\Bigl[-\bigl(\e^{ikr}+\overline{S_{\nu}(k)}\e^{-ikr}\bigr)\bigl(-ik^{\prime}\e^{-ik^{\prime}r}+ik^{\prime}S_{\nu}(k^{\prime})\e^{ik^{\prime}r}\bigr)\nonumber\\
    &\quad+\bigl(ik\e^{ikr}-ik\overline{S_{\nu}(k)}\e^{-ikr}\bigr)\bigl(\e^{-ik^{\prime}r}+S_{\nu}(k^{\prime})\e^{ik^{\prime}r}\bigr)\Bigr]\nonumber\\
    &=\lim_{r\to\infty}\left[\frac{\e^{i(k-k^{\prime})r}}{i(k-k^{\prime})}+\overline{S_{\nu}(k)}S_{\nu}(k^{\prime})\frac{\e^{-i(k-k^{\prime})r}}{-i(k-k^{\prime})}+S_{\nu}(k^{\prime})\frac{\e^{i(k+k^{\prime})r}}{i(k+k^{\prime})}+\overline{S_{\nu}(k)}\frac{\e^{-i(k+k^{\prime})r}}{-i(k+k^{\prime})}\right]\nonumber\\
    &=(1+\overline{S_{\nu}(k)}S_{\nu}(k^{\prime}))\pi\delta(k-k^{\prime})+(S_{\nu}(k^{\prime})+\overline{S_{\nu}(k)})\pi\delta(k+k^{\prime})\nonumber\\
    &=2\pi\delta(k-k^{\prime})+2\pi S_{\nu}(k^{\prime})\delta(k+k^{\prime}),\label{eq:A.7}
  \end{align}
  where in the first equality we have used the fact that the surface
  term at $r=0$ vanishes under the boundary condition
  \eqref{eq:20}. The second equality follows from the asymptotic
  behavior $\psi_{k}(r)\to\e^{-ikr}+S_{\nu}(k)\e^{ikr}$ as
  $r\to\infty$ and the fourth equality follows from the following
  formula:\footnote{Let $f$ be an arbitrary function. Then we have
    \begin{align}
      \int_{-\infty}^{\infty}\!\!dk\,f(k)\lim_{r\to\infty}\frac{\e^{ikr}}{ik}
      =\lim_{r\to\infty}\int_{-\infty}^{\infty}\!\!dz\,f(\tfrac{z}{r})\frac{\e^{iz}}{iz}
      =f(0)\int_{-\infty}^{\infty}\!\!dz\,\frac{\e^{iz}}{iz}
      =\pi f(0),\nonumber
    \end{align}
    where the integral should be understood as the Cauchy's principal
    value. Hence $\lim_{r\to\infty}\e^{ikr}/(ik)$ plays the role of
    $\pi\delta(k)$.}
  \begin{align}
    \lim_{r\to\infty}\frac{\e^{ikr}}{ik}=\pi\delta(k).\label{eq:A.8}
  \end{align}
  The last line of \eqref{eq:A.7} follows from \eqref{eq:34a} and
  \eqref{eq:34b}. Note that the $\delta$-function
  $\delta(k+k^{\prime})$ vanishes because $k+k^{\prime}$ never becomes
  zero for $k,k^{\prime}>0$. Hence we obtain \eqref{eq:35b}. This
  completes the proof of orthonormality.

  \subsection{Completeness}
  \label{appendix:A.2}
  Let us next prove the completeness \eqref{eq:36}. To this end, we
  first calculate the integral
  $\int_{0}^{\infty}\!\frac{dk}{2\pi}\,\psi_{k}(r)\overline{\psi_{k}(r^{\prime})}$. Substituting
  \eqref{eq:29} into this, we have
  \begin{align}
    \int_{0}^{\infty}\!\!\frac{dk}{2\pi}\,\psi_{k}(r)\overline{\psi_{k}(r^{\prime})}
    &=\int_{0}^{\infty}\!\!\frac{dk}{2\pi}\,(f_{\nu}(-kr)+S_{\nu}(k)f_{\nu}(kr))(\overline{f_{\nu}(-kr^{\prime})}+\overline{S_{\nu}(k)}\,\overline{f_{\nu}(kr^{\prime})})\nonumber\\
    &=\int_{0}^{\infty}\!\!\frac{dk}{2\pi}\,(f_{\nu}(-kr)+S_{\nu}(k)f_{\nu}(kr))(f_{\nu}(kr^{\prime})+S_{\nu}(-k)f_{\nu}(-kr^{\prime}))\nonumber\\
    &=\int_{0}^{\infty}\!\!\frac{dk}{2\pi}\,(f_{\nu}(-kr)f_{\nu}(kr^{\prime})+f_{\nu}(kr)f_{\nu}(-kr^{\prime})\nonumber\\
    &\quad+S_{\nu}(k)f_{\nu}(kr)f_{\nu}(kr^{\prime})+S_{\nu}(-k)f_{\nu}(-kr)f_{\nu}(-kr^{\prime}))\nonumber\\
    &=\int_{-\infty}^{\infty}\!\frac{dk}{2\pi}\,(f_{\nu}(-kr)f_{\nu}(kr^{\prime})+S_{\nu}(k)f_{\nu}(kr)f_{\nu}(kr^{\prime})),\label{eq:A.9}
  \end{align}
  where the second equality follows from \eqref{eq:22a} and
  \eqref{eq:34b} and the third equality follows from \eqref{eq:34a}
  and \eqref{eq:34b}.

  The last line of \eqref{eq:A.9} can be evaluated by just using the
  asymptotic behavior of the Jost-like solution $f_{\nu}$ and the
  simple-pole structure of the S-matrix $S_{\nu}$. First, we note that
  the product $f_{\nu}(-kr)f_{\nu}(kr^{\prime})$ behaves as
  $f_{\nu}(-kr)f_{\nu}(kr^{\prime})\to\e^{-ik(r-r^{\prime})}$ as
  $|k|\to\infty$. Hence, its integral along the infinite-radius upper
  semi-circle $C$ in the complex $k$-plane vanishes for
  $r-r^{\prime}<0$. (For $r-r^{\prime}>0$, one should rewrite the
  first term in \eqref{eq:A.9} as
  $\int_{-\infty}^{\infty}\frac{dk}{2\pi}f_{\nu}(-kr)f_{\nu}(kr^{\prime})=\int_{-\infty}^{\infty}\frac{dk}{2\pi}f_{\nu}(kr)f_{\nu}(-kr^{\prime})$,
  whose integrand vanishes along $C$ as well. Hence, without loss of
  generality we can focus on the case $r-r^{\prime}<0$.) Similarly,
  the product $S_{\nu}(k)f_{\nu}(kr)f_{\nu}(kr^{\prime})$ behaves as
  $S_{\nu}(k)f_{\nu}(kr)f_{\nu}(kr^{\prime})\to
  S_{\nu}(\infty)\e^{ik(r+r^{\prime})}$ as $|k|\to\infty$. Hence, its
  integral along the upper semi-circle $C$ vanishes as well. Thus the
  residue theorem gives
  \begin{subequations}
    \begin{align}
      \int_{-\infty}^{\infty}\!\frac{dk}{2\pi}\,f_{\nu}(-kr)f_{\nu}(kr^{\prime})+\int_{C}\frac{dk}{2\pi}\,f_{\nu}(-kr)f_{\nu}(kr^{\prime})&=0,\label{eq:A.10a}\\
      \int_{-\infty}^{\infty}\!\frac{dk}{2\pi}\,S_{\nu}(k)f_{\nu}(kr)f_{\nu}(kr^{\prime})+\int_{C}\frac{dk}{2\pi}\,S_{\nu}(k)f_{\nu}(kr)f_{\nu}(kr^{\prime})&=i\Res_{k=i\kappa_{0}}\left(S_{\nu}(k)f_{\nu}(kr)f_{\nu}(kr^{\prime})\right),\label{eq:A.10b}
    \end{align}
  \end{subequations}
  where we have used the fact that, for $g>0$, the S-matrix has only
  one simple pole at $k=i\kappa_{0}$ on the upper-half of the first
  Riemann sheet (see figure \ref{figure:2}). Writing $k$ as the polar
  form $k=R\e^{i\theta}$, we get
  \begin{align}
    \int_{-\infty}^{\infty}\!\frac{dk}{2\pi}\,f_{\nu}(-kr)f_{\nu}(kr^{\prime})
    &=-\int_{C}\frac{dk}{2\pi}\,f_{\nu}(-kr)f_{\nu}(kr^{\prime})\nonumber\\
    &=-\lim_{R\to\infty}\int_{0}^{\pi}\!\frac{iR\e^{i\theta}d\theta}{2\pi}f_{\nu}(-R\e^{i\theta}r)f_{\nu}(R\e^{i\theta}r^{\prime})\nonumber\\
    &=\lim_{R\to\infty}\int_{0}^{\pi}\!\frac{R\e^{i\theta}d\theta}{2\pi i}\e^{-iR(r-r^{\prime})\e^{i\theta}}\nonumber\\
    &=\lim_{R\to\infty}\left[\frac{1}{2\pi i}\frac{\e^{-iR(r-r^{\prime})\e^{i\theta}}}{r-r^{\prime}}\right]_{\theta=0}^{\theta=\pi}\nonumber\\
    &=\lim_{R\to\infty}\left(\frac{1}{2\pi i}\frac{\e^{+iR(r-r^{\prime})}-\e^{-iR(r-r^{\prime})}}{r-r^{\prime}}\right)\nonumber\\
    &=\lim_{R\to\infty}\int_{-R}^{R}\frac{dk}{2\pi}\e^{ik(r-r^{\prime})}\nonumber\\
    &=\int_{-\infty}^{\infty}\frac{dk}{2\pi}\e^{ik(r-r^{\prime})}\nonumber\\
    &=\delta(r-r^{\prime}),\label{eq:A.11}
  \end{align}
  where the third line follows from the asymptotic behavior
  \eqref{eq:22b}. Similarly, we have
  \begin{align}
    \int_{-\infty}^{\infty}\!\frac{dk}{2\pi}\,S_{\nu}(k)f_{\nu}(kr)f_{\nu}(kr^{\prime})
    &=i\Res_{k=i\kappa_{0}}\left(S_{\nu}(k)f_{\nu}(kr)f_{\nu}(kr^{\prime})\right)-\int_{C}\frac{dk}{2\pi}\,S_{\nu}(k)f_{\nu}(kr)f_{\nu}(kr^{\prime})\nonumber\\
    &=i\cdot i|N_{\kappa_{0}}|^{2}f_{\nu}(i\kappa_{0}r)f_{\nu}(i\kappa_{0}r^{\prime})+S_{\nu}(\infty)\delta(r+r^{\prime})\nonumber\\
    &=-\psi_{\kappa_{0}}(r)\overline{\psi_{\kappa_{0}}(r^{\prime})}+S_{\nu}(\infty)\delta(r+r^{\prime}),\label{eq:A.12}
  \end{align}
  where the last line follows from
  $\overline{f_{\nu}(i\kappa_{0}r^{\prime})}=f_{\nu}(i\kappa_{0}r^{\prime})$. Substituting
  \eqref{eq:A.11} and \eqref{eq:A.12} into \eqref{eq:A.9}, we get
  \begin{align}
    \psi_{\kappa_{0}}(r)\overline{\psi_{\kappa_{0}}(r^{\prime})}+\int_{0}^{\infty}\!\!\frac{dk}{2\pi}\,\psi_{k}(r)\overline{\psi_{k}(r^{\prime})}=\delta(r-r^{\prime})+S_{\nu}(\infty)\delta(r+r^{\prime}).\label{eq:A.13}
  \end{align}
  Note that the $\delta$-function $\delta(r+r^{\prime})$ vanishes on
  the half line $\mathbb{R}_{+}$ because $r+r^{\prime}$ never becomes
  zero for $r,r^{\prime}>0$. Hence we get \eqref{eq:36}. This
  completes the proof of completeness.
\end{appendices}

\printbibliography[heading=bibintoc]
\end{document}